\documentclass{article}

\pdfoutput=1 

\usepackage[preprint,nonatbib]{nips_2018}

\usepackage[utf8]{inputenc}

\usepackage{caption,float}
\usepackage{url}
\usepackage{newtxmath}

\usepackage{color}
\usepackage{soul}
\usepackage{booktabs}
\usepackage{listings}
\usepackage{stmaryrd}
\usepackage{tikz}

\newcommand{\secref}[1]{Section~\ref{sec:#1}}
\newcommand{\lstref}[1]{Listing~\ref{lst:#1}}

\newcommand{\tabref}[1]{Table~\ref{tab:#1}}

\newcommand{\secsref}[2]{Sections~\ref{sec:#1} and \ref{sec:#2}}

\newcommand{\tabsref}[2]{Tables~\ref{tab:#1} and~\ref{tab:#2}}

\definecolor{green-light}{rgb}{0.2,0.5,0.2}
\definecolor{dark-blue}{cmyk}{1,0.58,0,0.21}

\newfloat{lstfloat}{h!tbp}{lop}
\floatname{lstfloat}{Listing}

\lstdefinelanguage{exh}{
  morekeywords={of,quotient,set,subsetof,choice,enum,counter,value,alias,when,forall,in,end,sum,mul},
  morekeywords={symmetric,antisymmetric,antisymmetry,requires,require,is,not,trigger,integer},
  sensitive=true,
  morecomment=[l]{//},
  morestring=[b]{"}
}

\newcommand{\exhident}[1]{\ttfamily\small \color{purple} \$#1}

\newcommand{\bbrackets}[1]{\left\llbracket #1 \right\rrbracket}
\newcommand{\braces}[1]{\left\{ #1 \right\}}
\newcommand{\parens}[1]{\left( #1 \right)}
\newcommand{\scinum}[2]{#1\mathrm{e}#2}

\def\objects{\Omega}
\def\properties{\Pi}
\def\domains{\mathcal D}
\def\naturals{\mathbb N}
\def\kernel{X}
\def\choice{\chi}

\newcommand{\partsof}[1]{\mathcal P\parens{#1}}

\begin{document}

\title{On the Representation of Partially Specified Implementations and its Application to the Optimization of Linear Algebra Kernels on GPU}

\author{Ulysse Beaugnon\\
  \footnotesize École Normale Supérieure and Google\\
  \footnotesize \url{ulysse.beaugnon@ens.fr}\\
  \And
  Basile Clément\\
  \footnotesize Inria and École Normale Supérieure\\
  \footnotesize \url{basile.clement@ens.fr}\\
  \And
  Nicolas Tollenaere\\
  \footnotesize Inria and École Normale Supérieure\\
  \footnotesize \url{nicolas.tollenaere@inria.fr}\\
  \And
  Albert Cohen\\
  \footnotesize Google\\
  \footnotesize \url{4c0h3n@gmail.com}%
}

\maketitle

\begin{abstract}
  Traditional optimizing compilers rely on rewrite rules to
  iteratively apply program transformations.  This iterative approach
  hides optimization opportunities behind intermediate transformation
  steps.  For instance, vectorization can only be applied to the
  innermost loop in a nest: one must first perform a loop interchange
  before even considering vectorization of an outer loop.

  In contrast, we propose an implementation framework representing
  programs as sets of possible implementation decisions.  Specifying
  one decision can have an impact on others in a bidirectional manner:
  specifying that a loop must be vectorized prevents other loops from
  being nested inside it; conversely, specifying a loop as an outer
  loop will prevent it from being vectorized.  These optimization
  decisions commute, obviating the pass ordering problem.  We present
  a constraint programming system to formally define, represent and
  explore such implementation spaces.  We also propose an exploration
  strategy combining tree search and branch-and-bound; the strength
  and novelty of this strategy reside in an analytical model of the
  lower bound on the execution time of a set of possible
  implementations.

  We showcase our approach on the construction and exploration of an
  implementation space for linear algebra kernels running on GPUs.  We
  show this search space is expressive enough to represent complex
  decisions that fundamentally change the structure of the generated
  code.  We also present preliminary results competitive with the
  performance of native GPU libraries.
\end{abstract}

\section{Introduction}

General-purpose multicore processors encounter stiff competition from
specialized hardware accelerators such as Graphic Processing Units (GPUs).
Those accelerators usually offer massive parallelism, providing a raw
processing power orders of magnitude above that of general-purpose
processors. In some domains, adequately taking advantage of
this parallelism can be game changing; for instance, the recent success of deep
learning methods can largely be attributed to the performance of
GPU-accelerated libraries \cite{NIPS2012_4824, cuDNN}.

Unfortunately, writing or generating efficient code for GPUs is a complex
problem. It requires making careful decisions such as mapping computations to
the appropriate level of parallelism, moving data across the memory hierarchy
and through memory spaces of different sizes and properties, in addition to the
many possible thread-local optimizations.  While the set of potential
implementation decisions is usually well known, complex interactions between
them make it hard to find the best global implementation.  Taking one decision
can alter the low-level structure of the code or change the pressure on
constrained resources (such as available memory or hardware threads), which can
render other decisions invalid or inefficient. These interactions are critical
for performance; for instance, parallelizing too many loops can be harmful if
the threads end up doing too few computations.

The problem of picking a good implementation in an interaction-heavy
optimization space is not unique to GPUs and occurs on any architecture complex
enough that exhaustive enumeration is impossible.  Multiple approaches have
thus already been proposed to represent and explore the space of potential
implementations, such as SPIRAL \cite{spiral} for DSP codes, 
LGen \cite{lgen} for basic linear algebra, or LIFT \cite{lift} for more general computations expressed using combinators.
These approaches all share a reliance on rewrite rules to iteratively apply
transformations to the code. We argue that rewrite rules make it hard to anticipate possible downstream transformations and to derive
profitability information from one point in the search space: while the
transformations that can be directly applied are known, further ones may only become
available after so-called ``enabling'' transformations have been applied. Moreover, transformations
may not commute and different sequences of transformations may lead to the same
result. These factors make it hard to find the best implementation and as a
result, the production and the specialization to a given target or problem size
of highly tuned libraries remain a critical challenge.

\begin{itemize}
\item We propose an approach to formally define, represent and explore
  spaces of implementation candidates with partially instantiated
  decisions, and we model the interactions between such decisions as a
  constraint satisfaction problem.
\item Open decisions are listed explicitly and can be taken in any
  order, offering a well-behaved optimization space; in particular,
  the most performance-impacting decisions can be made first.
\item Precise information can be derived from sub-spaces even before
  all implementation decisions are made; this includes performance
  estimates and bounds.
\item We provide a method and tool to automatically generate efficient code to
  represent and manipulate partially specified implementations,
  starting from a high-level description; this tool could be applied
  to other classes of code optimization problems with different
  implementation decisions.
\item We demonstrate that the approach is expressive
  enough to represent complex decisions that fundamentally change the
  structure of the generated code, including compositions of
  strip-mining, loop fusion, loop interchange, unrolling,
  vectorization, mapping to multiple levels parallelism,
  and orchestrating data movements across the memory hierarchy.
\item We present preliminary results on the generation of optimized linear
  algebra kernels for GPUs, competitive with native libraries.
\end{itemize}

\secref{overview} explains how we represent implementation
candidates, \secref{exh} presents the framework we use to describe
potential decisions and their interactions and \secref{encoding} applies this
framework to linear algebra on GPUs.  Then \secref{results} shows its practical
benefits and \secref{discussion} compares our work with other approaches.

\section{Partially Specified Implementations}
\label{sec:overview}

Our first goal is to build a representation for implementation spaces that
is suitable for search algorithms.  We build this representation on the notion
of \emph{implementation candidate}---or \emph{candidate} for short---which is
a partially specified implementation.  A search algorithm works on
implementation candidates, which are further refined until they become fully
specified implementations that can be executed.  All candidates refer to a common \emph{semantic backbone} describing the computation to implement.

\subsection{Partially Instantiated Decision Vector}
\label{sec:partial-implem}

We represent implementation candidates as partially instantiated vector of
decisions that specify how to implement computations. Decisions are variables
of a Constraint Satisfaction Problem (CSP). Each variable has a small set of
values it can take, called its \emph{domain}. A \emph{constrained domain}
contains a single value.

The objective of a search algorithm is to find an assignment of variables 
respecting the constraints by successively restricting the domain
of each variable. The decision vector is the product of the domains and thus a
Cartesian over-approximation of the set of implementations. Two domains may
contain legal values that are incompatible with each other because of some
constraint.

A search algorithm typically builds a tree
whose nodes are candidates. The root is the fully unconstrained candidate.
From there, each node is a strictly more constrained version of its
parent. Fully constrained nodes have no children and correspond to
actual implementations for which the algorithm can generate code and measure
execution time.

After restricting a domain, constraint propagators must be run to remove
incompatible values from other domains. In the general case, some incompatible values
may not be detected before domains are further restricted, thus requiring
to backtrack. \secref{experiments:deadend} shows that the constraints we use
are simple enough that this is rarely needed in practice compared to usual
backtracking algorithms.

This approach offers two main advantages. First, domains define a well behaved set
of actions the algorithm can take to generate implementations. In
particular, the set of available actions decreases when descending in the tree,
and the order in which actions are selected does not matter. This allows the
algorithm to make the most performance-impacting decisions first.

Second, we can define functions on candidates to drive the search.
For instance, we can compute an estimate of the execution time or a probability
distribution on the decisions to take next.  
These functions can derive relevant information about the sets of potential
implementations described by the domains instead of reasoning on a single
implementation at a time.

\subsection{Constraint Satisfaction Problem}
\label{sec:csp-def}

The decision vector alone is not enough to generate code. It does not specify
the semantics of the code to generate nor how to instantiate decisions for
different kernels.  This information is present in a separate structure describing the computation to implement.
We call this structure a \emph{semantic backbone}; it is mostly
constant through the search, although some decisions such as introducing temporary variables can interact in a limited way with the backbone.  Decisions can be understood as defining properties
on the objects in the backbone: for instance, if the backbone defines a loop, it induces a decision for unrolling the loop.

Let us now present the formalism to derive a CSP
from a backbone instance.  It describes generic choices and constraints to
instantiate on each semantic backbone to define the decision vector. For that,
we view the backbone as a set of objects that respect different properties. For
example, properties can be \emph{$x$ is an instruction} or
\emph{$x$ depends on $y$}.
While objects vary from one kernel to the other, properties
remain the same. They allow to describe the implementation space independently
of the backbone instance.

Formally, let $\objects_0$ be a set of identifiers representing basic objects.
We define the set of objects $\objects$ as the set of nested tuples of basic
objects: for $n > 0$, we recursively define $\objects_{n + 1} = \objects_n \cup
\objects_n^*$ where $\objects_n^*$ is the set of tuples of arbitrary length,
and $\objects = \cup_{n \in \naturals} \objects_n$.  Note that this definition
implies $\objects^k \subseteq \objects$ for all $k$.

Let $\properties$ be a set of identifiers called \textit{properties};
$a: \properties \to \naturals$ denotes their arity.  A
backbone instance is a function $\kernel$ mapping properties to the
objects satisfying it:
\[
  \begin{array}{ll}
    \kernel: & \properties \to \partsof{\objects} \\
	     & p \mapsto \kernel(p) \subseteq \objects^{a(p)}
  \end{array}
\]

Finally, let $\domains$ be a set of \textit{domains}, where each domain is a
(distinct) set of identifiers. A generic choice is a tuple $c: ((p_0, \dots
p_{k-1}), D) \in \properties^* \times \domains$. It indicates a decision among
the values of $D$ for each tuple of objects verifying the properties.  For
instance, the decision could be how each loop should be implemented (regular
loop, unrolled loop, parallel loop, etc.) or scheduling information for pairs
of statements (recall that objects can be tuples of basic objects).

We define an \textit{implementation space} as the combination of a set $C
\subseteq \properties^* \times \domains$ of generic choices, and a set of
first-order formulae called \textit{constraints}.  The alphabet for the
constraints contains one predicate symbol with arity $a(p)$ for each property
$p \in \properties$, and one function symbol with arity $a(p_0) \times \cdots
\times a(p_{k-1})$ for each choice $c = ((p_0, \ldots, p_{k-1}), D) \in C$.  

From there, a \textit{partially instanciated decision} for a choice $c = ((p_0, \ldots,
p_{k-1}), D)$ is a function:
\[
	\choice_c: X(p_0) \times \cdots \times X(p_{k-1}) \mapsto \partsof{D}.
\]

The decision is partial in the sense that it restricts the possible values, but
does not necessarily select one.  A partially instantiated decision is
\textit{invalid} if its target domain contains the empty set, and is
\textit{fully instanciated} if its target domain only contains singletons.

For a given implementation space and a backbone $\kernel$, a \textit{decision
vector} or \textit{implementation candidate} (\textit{candidate} for short) is
a vector with one partially instantiated decision for each choice.  An
\textit{implementation} is an implementation candidate where
\begin{itemize} 
\item all decisions are fully instantiated;
\item all the implementation space constraints hold when interpreting
  the predicate symbol for property $p$ with $\kernel(p)$ and the
  function symbol for choice $c$ with $\choice_c$.
\end{itemize}

This formalism allows us to dynamically extend the implementation space
during search. Extending the set of objects respecting properties adds
decisions and constraints without invalidating pre-existing ones. This is
useful to lower specific constructs upon reaching decisions opening for further refinement of the implementation choices. For example, allowing to
insert the appropriate load and store instructions after deciding to allocate a temporary array in memory. Lowerings may only add new objects or add properties to existing ones.

It is important to note that such lowerings only concern the CSP solver. We
pre-compute all possible lowerings ahead of time for search heuristics.
Lowerings allow to delay introducing variables and avoid cluttering the CSP
with decisions and constraints that may not affect the generated code.
The same properties can always be encoded with regular constraints instead,
ensuring this does not contradict our claim that all potential decisions
can be available upfront.

\section{Implementation Space Description}
\label{sec:exh}

We now present the details of the search space definition framework following
the formalism from \secref{csp-def}. It exposes a high-level
language to define the different properties exposed by the semantic backbone
(\secref{exh:sets}), the decisions (\secref{exh:choices}) and the constraints
(\secref{exh:constraints}). It also specifies when to add new objects to the
semantic backbone (\secref{exh:lowering}). A compiler reads this language and
generates code to create, represent and manipulate the partially instantiated
vector of decisions. In particular, it generates code to propagate the
constraints, inserts new objects in the backbone when needed and extend the CSP
with the corresponding new choices and constraints.
This framework allowed us to experiment with multiple encodings for linear
algebra kernels on GPUs with limited effort.  The same framework can easily be
reused for different application domains or hardware targets.

\subsection{Interface With the Kernel Representation} \label{sec:exh:sets}

The first step to specify an implementation space is to define sets of objects
exposed by the semantic backbone. Each set corresponds to a property $p \in
\properties$ and is instantiated into $\kernel(p)$ for each kernel instance
$\kernel$.  Definitions indicates both the relation of the set to others and
how to generate code that manipulates its objects. This way, we can easily apply
our approach to kernel representations. Exposing a new property requires less
than 10 lines of code.

\begin{lstfloat}[htp]
  \begin{minipage}[b]{.48\textwidth}
\begin{lstlisting}[language=Exh,caption={Set Definitions},label=lst:exh:set-def]
set Instructions:
  type = "Instruction"
  iterator = ...
  item_getter = ...
  ... // Elided for brevity
end

set MemAccesses subsetof Instructions:
  type = "MemAccess"
  ... // Elided for brevity
end

set AccessedRegions(!\exhident{inst}! in MemAccess):
  ... // Elided for brevity
end
\end{lstlisting}
\end{minipage}
\hfill
  \begin{minipage}[b]{.48\textwidth}
\begin{lstlisting}[language=exh,
                   caption={\hbox{Choice Definition for Cache Directives}},
                   label=lst:exh:simple-enum]
choice enum cache(!\exhident{inst}! in MemAccesses):
  value L1:        // Use L1+L2 caches
  value L2:        // Use L2 cache
  value READ_ONLY: // Use read-only cache
  value NONE:      // Do not use caches
end
\end{lstlisting}

\begin{lstlisting}[language=exh,
                   caption={\hbox{Choice Definition for Ordering Decisions}},
                   label=lst:exh:order]
choice enum order(!\exhident{lhs}! in Instructions,
                  !\exhident{rhs}! in Instructions):
  value BEFORE:
  value AFTER:
  antisymmetric:
    BEFORE -> AFTER
end
\end{lstlisting}
\end{minipage}
\end{lstfloat}

\begin{lstfloat}
\begin{lstlisting}[language=exh,
                   label=lst:exh:order-transitive,
                   caption={Order Transitivity Constraint}]
require forall !\exhident{a}! in Instructions:
  forall !\exhident{b}! in Instructions:
    forall !\exhident{c}! in Instructions:
      order(!\exhident{a}!, !\exhident{c}!) is AFTER
      || order(!\exhident{a}!, !\exhident{b}!) is not AFTER
      || order(!\exhident{b}!, !\exhident{c}!) is not AFTER
\end{lstlisting}

\begin{lstlisting}[language=exh,
                   caption={Local Memory Usage Counter},
                   label=lst:exh:counter]
choice counter local_mem():
  forall !\exhident{mem\_block}! in MemBlocks:
    sum "!\exhident{mem\_block}!.size()" when
      mem_space(!\exhident{mem\_block}!) is LOCAL
end

require local_mem() < "gpu.local_mem_size"
\end{lstlisting}

\begin{lstlisting}[
  language=exh,
  label=lst:exh:quotient,
  caption=Thread Dimensions Quotient Set
]
choice enum fused(!\exhident{lhs}! in Dimensions,
                  !\exhident{rhs}! in Dimensions):
  ... // Elided
end

quotient ThreadDims of !\exhident{dim}! in Dimensions:
  is_thread_dim = dim_kind(!\exhident{dim}!) is THREAD
    / fused is TRUE
  ... // Elided
end

choice counter num_threads():
  forall !\exhident{dim}! in Dimensions:
    sum size(!\exhident{dim}!) when:
      is_thread_dim(!\exhident{dim}!) is TRUE
end
\end{lstlisting}
\end{lstfloat}

For example, \lstref{exh:set-def} defines a set \texttt{Instructions} that contains
objects of type \texttt{Instruction} in the host language. The programmer specifies fields
to indicate how to iterate on the set and to retrieve its objects.

Programmers can express inclusion and disjointness relations between sets. They allow to
declare a set structure that matches the subtyping relation in the host language. For
instance, \texttt{MemAccesses} is a subset of \texttt{Instructions} and contains objects
of type \texttt{MemAccess}, a subtype of \texttt{Instruction}.

Last we can parametrize sets with another set. For example, \lstref{exh:set-def} defines
an \texttt{AccessedRegions} set for each memory access. This translates to a binary property
containing tuples of \texttt{AccessedRegions x MemAccess} in our formalism.

\subsection{Choice Definition} \label{sec:exh:choices}

Next we define the set of generic choices $C$. Each $c \in C$ has a list of sets
corresponding to properties $p_0, \dots p_{k-1} \in \properties$ and a domain $D_c \in
\domains$. When instantiated for a particular kernel $\kernel$, the choice defines a function
$\choice_c$ from tuples of distinct objects in $\kernel(p_0) \times \kernel{(p_{k-1})}$ to $D_c$.

For example, \lstref{exh:simple-enum} defines a function $cache$ from memory accesses to
cache directives:
\[cache: \texttt{MemAccesses} \to
  \braces{\texttt{L1},\, \texttt{L2},\, \texttt{READ\_ONLY},\, \texttt{NONE}}\]
This defines a CSP variable for each memory access exposed by the semantic backbone.  The
goal of the implementation space exploration is then to assign a value $v \in D_c$ to
$\choice_c(x_0, \dots x_{n-1})$ for each $\parens{x_0, \dots x_{n-1}} \in \kernel(p_0) \times
\dots \times \kernel\parens{p_{n-1}}$ with $x_0, \dots x_{n-1}$ distinct. In our example,
this is assigning a cache directive to each memory access.

Another example, \lstref{exh:order} defines a generic choice $order$ from pairs of
instructions to orderings.
It first illustrates why we only consider tuples of distinct objects: it is the behavior
expected by the programmer when defining choices. The order between an instruction and
itself would not make sense. It also illustrates how we define antisymmetric choices,
allowing the code generator to generate stronger propagators and only store half of the
domains of \verb!order! variables.

We currently support two kinds of domains:
\begin{itemize}
  \item \verb!enum! choices, such as \verb!cache! in \lstref{exh:simple-enum}, that can
    take a small set of predefined values and
  \item \verb!integer! choices, that take values in a small universe specific to each
    instance of the choice. The choice definition includes a piece of code that retrieves
    the universe from the kernel representation.
\end{itemize}

Our tool automatically generates types to represent domains and code to instantiate the
generic choices for a semantic backbone instance.

\subsection{Constraints} \label{sec:exh:constraints}

Next, we express constraints on variables domains to avoid invalid implementations.
We use constraints to:
\begin{itemize}
\item avoid incoherent decisions, e.g., non-transitive ordering of instructions;
\item enforce correctness constraints imposed by the semantic backbone, e.g., data dependencies;
\item and respect hardware-specific limitations, e.g., the size of the memory blocks allocated to a memory space cannot exceed the size of that memory space.
\end{itemize}

Constraints are first order logic sentences on the choices, quantified over objects in specific sets. Currently, they are universally quantified disjunctions of conditions on zero, one or two choices: boolean constants, restrictions of
a variable to specific values or comparisons between two variables. For
example, \lstref{exh:order-transitive} ensure the \verb!order! choice defined
in \lstref{exh:order} is transitive with a constraint applied to all triples of
distinct instructions \verb!a!, \verb!b! and \verb!c!.

Constraints implicitly assume quantified objects are distinct. This makes it easier to write concise constraints and matches the intended behavior in most cases. Our tool automatically inserts clauses to enforce this in the generated code.

Constants can be pieces of code parametrized by the objects.
This enables us to parametrize constraints at a fine grain without having to
declare new sets. In particular, it allows to:
\begin{itemize}
  \item parametrize bounds on integer decisions
    \begin{lstlisting}[language=exh]
size(!\exhident{mem}!) <= "!\exhident{mem}!.max_size"
    \end{lstlisting}
  \item selectively disable constraints for some objects:
    \begin{lstlisting}[language=exh]
"!\exhident{inst}!.is_vectorizable" || <constraint>
    \end{lstlisting}
\end{itemize}

Our tool generates code to propagate constraints. It restricts domains
to remove incompatible decisions at the initialization of the implementation space and
after restricting any domain. The simple form of our constraints allow to easily generate
efficient propagation code. For fixed object variables, a constraint only reference a few
choices. This limits the amount of propagation to perform after each update.

Often, hardware limitations impose constraints on sums or products of quantities, such as
the sum of the memory region sizes (to ensure they fit in memory) or the product of the
size of nested thread dimensions (to ensure we do not exceed the maximal number of
threads). Simple constraints cannot express such values. Instead, one can define
\emph{counters} that track sums or products of values.
For example, \lstref{exh:counter} keeps track of the local
memory usage.

While remaining simple, our constraints are expressive enough to encode correct implementations. Because the propagation code is automatically generated, it is easy to experiment with and encode a wide range of optimization decisions.

\subsection{Kernel Representation Lowering} \label{sec:exh:lowering}

Finally, we define \emph{triggers} to lower backbone constructs when certain conditions are met.
Triggers run a callback that either adds properties to existing objects or
creates new objects. For example, it can add the property \emph{is stored in
memory} to a variable, creating new decisions to specify the memory layout.
At the same time, the trigger can also add a store and a load
instruction to refine placement decisions in the memory hierarcy.

Callbacks may only add objects to properties, not remove them so that previous variables
and constraints remain valid. Moreover, we require that callbacks commute. It is important to
understand that triggers are not mandatory: one can add objects to properties upfront and
condition decisions and constraints with the lowering condition. In our example, cache
directives would be forced to \texttt{None} in candidates where memory accesses do not
exist. The reasons to use triggers are:
\begin{itemize}
\item to avoid cluttering the decision vector with decisions that will not affect the generated code, making constraint propagation more efficient;
\item and intentionally delay some decisions to focus the search algorithms on more meaningful choices first.
\end{itemize}

As with constraints, we define lowering conditions as first order sentences,
with universally quantified object variables. This allows definitions to be
independent of backbone instances.

Triggers can be used to define quotient sets:
we often end-up needing to account for classes of equivalent objects
such as fused loops. Quotient sets contain one representative for each class of
objects respecting some condition. They are automatically maintained by our tool
using triggers. For example, in \lstref{exh:quotient} we count the number of threads.
This is done with a quotient set containing one representative for each class of fused
iteration dimensions mapped to a hardware thread. This also define a boolean choice,
\verb!is_thread_dim! indicating which dimensions are the representative of their class
(this is redundant, but helps when writing constraints).

Overall, we built a simple, domain-specific constraint programming framework
that facilitates the construction of search spaces, involving new choices and
constraints. The framework also eases porting our approach to different
architectures or application domains. Because we abstract the kernel
representation as sets of objects, the definition is independent of the kernel
instance. We could probably implement our tool on top of a standalone CSP
framework; apart from the need for a domain-specific front-end to provide the
above-mentioned kernel genericity, this would involve several optimizations and
customizations to reproduce the search strategy and match the efficiency of our
approach, motivating a custom design.
In particular, we use the fact that the CSP definition is independent of the backbone
to avoid storing constraints instances in memory. Instead we statically
generate code that we call for different objects. This reduces the memory
footprint of the CSP to the sole decision vector, making it compact, easy to serialize
for logging and fast to copy.

\section{Partial Implementations for GPUs}
\label{sec:encoding}
This section builds on the framework presented in \secref{exh} representing partial
implementations for linear algebra kernels on GPU. It serves both as an
illustration of our approach and as an effective way to generate competitive code on GPU.

Since this representation targets dense linear algebra, we assume all loops are
\emph{for}-loops with data-independent bounds (i.e., loop bounds only depend on the input
size) and affine memory accesses to multidimensional tensors. We focus on
optimizations that work well on GPUs. Other representations can be developped with
similar ideas using the framework presented in \secref{exh}.

We highlight the set and choice definitions and omit counters and
constraints for the sake of brevity. We highlight the main
difficulties and original aspects of code generation, deferring more
systematic coverage to a later, dedicated paper.

\subsection{Statements}

We represent computations to perform as a list of instructions to execute.
Each instruction performs a single scalar operation: an arithmetic operation (addition,
multiplication, cast, \dots) or a memory access (load or store). We then specify
how many time instructions are executed by nesting them into iteration dimensions.
Iteration dimensions are akin to counted for-loops but may be implemented differently.
Depending on decisions, they can be unrolled, vectorized or parallelized.

\begin{table}[ht]
  \centering
  \begin{tabular}{ll}
    \toprule
    Instructions & \text{Dimensions} \\
    \midrule
    \verb!a = load A[i_m]! & $\braces{d_m}$\\
    \verb!b = load B[i_n]! & $\braces{d_n}$\\
    \verb!c = a * b! & $\braces{d_m, d_n}$ \\
    \verb!store C[i_m * n + i_n] <- c! & $\braces{d_m, d_n}$ \\
    \bottomrule
  \end{tabular}
  \caption{Outer Product of Two Vectors Kernel}
  \label{tab:encoding:mat-mul-insts}
\end{table}

\tabref{encoding:mat-mul-insts} shows an example of kernel that computes the outer
product of two vectors $A$ and $B$ of sizes $m$ and $n$ and stores the result in a matrix
$C$ of size $m \times n$. This kernel has two iteration dimensions $d_m$ and
$d_n$ of respective size $m$ and $n$ and current index $i_m$ and $i_n$.
Note that this representation does not imply a particular order of instructions
or nesting of dimensions.

\begin{lstfloat}[htp]
\begin{lstlisting}[
  language=exh,
  caption={Control-Flow Related Defintions},
  label=lst:encoding:statements
]
set Statements: ...
set Insts subsetof Statements: ...
set Dimensions subsetof Statements: ...

choice enum order(!\exhident{lhs}! in Statements,
                  !\exhident{rhs}! in Statements):
  // $lhs is executed before $rhs
  value BEFORE:
  // $rhs is executed before $lhs
  value AFTER:
  // $lrhs is nested inside $rhs
  value INNER:
  // $rhs is nested inside $lhs
  value OUTER:
  // $lhs and $rhs are fused dimensions
  value MERGED:
  antisymmetric:
    BEFORE -> AFTER
    INNER -> OUTER
end

quotient IterationDims(!\exhident{inst}! in Insts)
of !\exhident{dim}! Dimensions:
  is_outer_dim = order(!\exhident{dim}!, !\exhident{inst}!) is
    OUTER / order is MERGED
  ...
end
\end{lstlisting}
\end{lstfloat}

\lstref{encoding:statements} shows the definition of two sets listing the instructions and
the dimensions. Together, they form the statements. The \verb!order! choice
dictates the control flow structure. It simultaneously encodes statements sequential
ordering (\verb!BEFORE! and \verb!AFTER!), nesting (\verb!INNER! and
\verb!OUTER!), and loops fusion (\verb!MERGED!). In particular, it can:
\begin{itemize}
  \item move code inside or outside of loops, for example by setting
    \verb!order($stmt, $dim)!  to \verb!BEFORE! or \verb!INNER!,
  \item interchange loops, by setting their order to \verb!INNER! or \verb!OUTER!.
  \item fuse loops by setting their order to \verb!MERGED!.
  \item schedule loops and instructions by setting the order between two statements to
    \verb!BEFORE! or \verb!AFTER!.
\end{itemize}

The \verb!order! choice demonstrates the first benefit of our approach: the
compiler is not limited to a set of predefined high-level transformations. It
does not even have to be aware of them.  Instead we expose many smaller decisions
(here, the pairwise ordering between statements). High level transformations
are implemented as specific combinations of decisions.  We are free to pick any other
assignment that respects the constraints.  This approach makes it easier to
understand the interaction between transformations (as decisions) and to combine them since they are
all exposed in the same framework.

\lstref{encoding:statements} also defines a quotient set \verb!IterationDims! that
contains the iteration dimensions nested outside each instruction. At the
creation of the domain, it specifies the nesting imposed by the backbone.
Afterward, new dimensions may be added when \verb!order! gets constrained. It
is grouped into equivalence classes by the \texttt{order is MERGED} relation so
that fused dimensions are only accounted for once.

\begin{lstfloat}[htp]
\begin{lstlisting}[
  language=python,
  caption={Nestings for the Outer Product Kernel},
  label=lst:encoding:impossible-order,
]
# Either d_n and `b` are nested inside d_m
for i in 0..M:   # d_m
  a = load A[i]
  for j in 0..n: # d_n
    b = load B[j]
    c = a * b
    store C[i*M+j] <- c
# or d_m and `a` are nested inside d_n.
for j in 0..N:   # d_n
  b = load B[j]
  for i in 0..N: # d_m
    a = load A[i]
    c = a * b
    store C[i*M+j] <- c
\end{lstlisting}
\end{lstfloat}

Together with \verb!order!, \verb!IterationDims! demonstrates another benefit of our
approach: the structure of the code does not needs to be coherent before the candidate
is fully constrained. For example, in \tabref{encoding:mat-mul-insts}, \verb!load A! is
nested in $d_m$, \verb!load B! in $d_n$ and $c = a * b$ in $d_m$ and $d_n$. As shown in
\lstref{encoding:impossible-order}, such a loop structure is impossible: either
$d_n$ and \verb!load B! are nested inside $d_m$ or $d_m$ and \verb!load A! inside $d_n$.
Here it is legal because ordering decisions are still open. We can delay choosing the
nesting while a regular control flow representation would need to pick one.

\subsection{Iteration Dimensions}

The \verb!dim_kind! choice, shown in \lstref{encoding:dim-kind}, specifies how to
implement dimensions. They can be parallel (at the \verb!BLOCK! or \verb!THREAD!
levels), fully unrolled or imply vector instructions.

\begin{lstfloat}[htp]
\begin{lstlisting}[
  language=exh,
  caption={Encoding of Dimension Decisons},
  label=lst:encoding:dim-kind,
]
choice enum dim_kind(!\exhident{dim}! in Dimensions):
  // $dim is a regular for loop.
  value LOOP:
  // $dim is a hardware block dimension.
  value BLOCK:
  // $dim is a hardware thread dimension.
  value THREAD:
  // $dim is totally unrolled.
  value UNROLL:
  // !\color{green-light}\$!dim is vectorized.
  value VECTOR:
end
\end{lstlisting}
\end{lstfloat}

We usually strip-mine dimensions of the original code into multiple ones when creating
the search space. This allows both to apply different \verb!dim_kind! decisions to the
resulting dimensions and to tile computations to improve locality.

\begin{lstfloat}[htp]
\begin{lstlisting}[
  language=exh,
  caption={Encoding of Strip-Mining and Tiling},
  label=lst:encoding:dim-tiling,
]
set StaticDims subsetof Dimensions: ...

choice integer size(!\exhident{dim}! in StaticDims):
  "!\exhident{dim}!.possible_sizes()"
end

set LogicalDims: ...
set TilingDims(!\exhident{logical}! in LogicalDims)
  subsetof StaticDims: ...
set TiledDim(!\exhident{logical}! in LogicalDims)
  subsetof Dimensions: ...
\end{lstlisting}
\end{lstfloat}

\lstref{encoding:dim-tiling} shows how we encode strip-mining. We split
dimensions in two categories: \verb!StaticDims! that have a statically known
size, specified by the \verb!size! choice, and dynamic dimensions (not exposed
in a set) whose size depend on input parameters.
Logical dimensions represent dimensions of the original code. They are formed
of static dimensions, listed in \verb!TilingDims($logical_dim)! and zero or one
dynamic dimensions, listed in \verb!TiledDim($logical_dim)!, depending on whether the
logical dimension has a static or dynamic size.

\subsection{Memory Accesses}

\lstref{encoding:mem-insts} shows how we encode cache directives and memory placement.
\verb!MemRegions! are distinct pieces of memory in which we allocate arrays.
While memory regions holding input arrays are always in RAM, others regions may
also be placed in shared memory, local to a group of threads.

\begin{lstfloat}[htp]
\begin{lstlisting}[
  language=exh,
  caption={Encoding Memory Placement Decisions},
  label=lst:encoding:mem-insts
]
set MemRegions: ...
set MemInsts subsetof Insts: ...
set AccessedRegions(!\exhident{mem}! in MemInsts): ...

choice enum mem_space(!\exhident{mem}! in MemRegions):
  // $mem is stored in RAM.
  value GLOBAL:
  // $mem is stored in the memory shared
  // by a group of threads.
  value SHARED:
end

choice enum cache(!\exhident{inst}! in MemInsts): ...
\end{lstlisting}
\end{lstfloat}

\verb!MemInsts! lists loads and stores and \verb!AccessedRegions! the memory regions
accessed by such instructions. \verb!cache! indicates which caches to use (see
\lstref{exh:simple-enum} for the full definition).

\subsection{Instruction Operands}
\label{sec:encoding:operands}

Instruction operands are either constants, kernel inputs, induction variables or values
produced by instructions. Induction variables are linear combination of dimension
indexes. We use them to compute memory accesses addresses. We handle them
separately from regular instructions as it is clear how they should be implemented,
and exposing them in the search space would only increase complexity.

By default, operands can only take the last value produced by preceding instructions. To
implement reductions, an instruction operand can also take the value produced by the same
instruction at the previous iteration on a given set of dimensions. In that case, we
specify another instruction to initialize the reduction.

Additionally, we can specify point-to-point communication between dimensions: the value
produced at iteration $i$ of a dimension is consumed at iteration $i$ of another
dimension. For example, in \lstref{encoding:dim-map}, the instruction nested in \verb+d0_b+ and
\verb+d1_b+ reads the values produced in \verb+d0_a+ and \verb+d1_a+.

\begin{lstfloat}[htp]
\begin{lstlisting}[
  language=python,
  caption={Point-to-Point Communication Between Dimensions},
  label={lst:encoding:dim-map},
]
# Point-to-point communication
for i in 0..M:   # d0_a
  for j in 0..N: # d1_a
    x = load A[i][j]
for i in 0..M:   # d0_b
  for j in 0..N: # d1_b
    # y = 2 * A[i][j]
    y = 4 * x[d0_a -> d0_b, d1_a -> d1_b]

# Possible implementation.
for i in 0..M:   # fused d0_a and d0_b
  for j in 0..N: # d1_a
    x = load A[i][j]
    store TMP[j] <- x
  for j in 0..M: # d1_b
    x2 = load TMP[j]
    y = 4 * x2
\end{lstlisting}
\end{lstfloat}

Point-to-point communications allows putting each instruction in its own loop nest. This
is useful as it gives us more flexibility to schedule, fuse and implement dimensions.
Point-to-point communication is either implemented by:
\begin{itemize}
  \item fusing both dimensions;
  \item unrolling or vectorizing both dimensions, with the data stored in different
    registers for each iteration;
  \item mapping the two dimensions to the same hardware thread dimension, as
    explained in \secref{encoding:thread-mapping};
  \item or using a temporary array.
\end{itemize}

A trigger automatically creates a temporary array and the associated load and store
when other options are impossible. In particular, this allows copying
data to a temporary array in a faster memory for improving locality. For example
\lstref{encoding:dim-map} shows a possible implementation that stores chunks of \verb!A!
in an array \verb!TMP!.

\subsection{Thread Mapping}
\label{sec:encoding:thread-mapping}

GPUs expose two levels of parallelism: threads and blocks of threads. Thread
dimensions define a block of threads that share a fast memory and can
synchronise. Block dimensions replicate blocks in parallel. Each block or
thread dimension can be mapped to one of three hardware dimensions, hereafter
referred to as levels to distinguish them from iteration dimensions.

Blocks cannot easily communicate so block dimensions are outermost parallel
dimensions that span the entire computation. On the other hand,
synchronization within a thread block is critical to achieve good performance.
We encode synchronisation barriers by creating multiple loop nests of thread
dimensions that maps to the same hardware level. At code generation, we fuse
the loop nests and insert a synchronisation instruction between them.
In particular, this allows point-to-point communications between thread
dimensions mapped to the same level.

\begin{lstfloat}[htp]
\begin{lstlisting}[
  language=exh,
  caption={Thread Dimensions Mapping},
  label=lst:encoding:thread-mapping,
]
choice enum thread_level(!\exhident{x}! in StaticDims,
                         !\exhident{y}! in StaticDims):
  // maps $x and $y to the same level.
  value MAPPED:
  // maps $x to an inner level than $y.
  value INNER:
  // maps $y to an inner level than $x.
  value OUTER:
  // $x or $y is not a thread dimension.
  value NOT_THREADS:
  antisymmetric: INNER -> OUTER
end

quotient HwThreadDims of !\exhident{d}! in StaticDims:
  is_thread_dim = dim_kind(!\exhident{d}!)
    / thread_level is MAPPED
  ...
end
\end{lstlisting}
\end{lstfloat}

The \verb!thread-mapping! choice in \lstref{encoding:thread-mapping} specifies
how thread dimensions map to hardware levels. The nesting order of thread
dimension is crucial as it determines memory coalescing.
The number of threads may vary between thread nests: they might
not map to the same levels. In that case, we use predicated instruction to
disable some threads.

\section{Experiments}
\label{sec:results}

Let us now evaluate our approach on concrete search and code generation problems.
We implemented a search strategy that combines a statistical component
with a performance model of implementation candidates
(\secref{experiments:strategy}). Our goal is to show that:
\begin{itemize}
  \item we can generate code competitive with reference hand-tuned libraries
    (\secref{experiments:perf});
  \item the formalism and performance model allows to extract pertinent information
    before specifying decisions (\secref{experiments:info});
  \item and we improve search performance by making decisions commute
    (\secref{experiments:order}).
\end{itemize}

All experiments are run on a Linux machine equipped with a
12-core Xeon E5-2620v2 processor, 64GB of RAM and a Quadro K4000 GPU Kepler GPU
running under CUDA 8.0.

\subsection{Search Strategy}
\label{sec:experiments:strategy}


The search space exploration is driven by a Monte-Carlo Tree Search (MCTS)
algorithm.  We use a variant of Threshold Ascent on Graph (TAG), which was
previously applied to Spiral \cite{de_mesmay_bandit-based_2009}, a code
generator for fast Fourier transforms.

In our case, the algorithm builds a tree whose nodes are candidates.  It starts
with only the root node, representing the full search space. It iteratively
selects a leaf to expand by using the TAG formula and evaluation statistics from previous
iterations. The leaf expansion creates one child
per possible value of the decision. Then, the MCTS performs a Monte-Carlo simulation
to set remaining decisions. It runs the resulting implementation on the GPU and
adjusts the statistics along the selected path with the execution time.
The order in which decisions are taken is fixed upfront. We
manually select an order that specifies most important decisions
first. \secref{experiments:order} discusses the impact of the order on
exploration performance.


We complement the TAG algorithm with a performance model of the candidates
\cite{telamon_cc17}. The model provides a lower bound on the execution time of
all implementations derivable from a candidate. We use the bound in two ways:
\begin{itemize}
	\item In the selection phase, we ignore children which have a lower
		bound higher than the execution time of the current best
		implementation.
	\item During Monte-Carlo simulations, when choosing amongst candidates
		$(X_i)_i$, we pick $X_i$ with probability:
		\[
			p(X_i) \sim \max(T - b(X_i), \, 0)
		\]
		where $T$ is the execution time of the best implementation so
		far and $b(X_i)$ is the lower bound given by the performance
		model to the candidate $X_i$.  Note that $p(X_i) = 0$ when
		$b(X_i) > T$.
\end{itemize}

In both cases, the performance model prunes regions of the search space which
cannot possibly improve on the current best implementation found.  We show in
\secref{experiments:info} that this can eliminate large portions of the search
space.

Since this paper focuses on the definition, representation and
exploration of the optimization space, we defer deeper treatment and
analysis of the seartch strategy---involving MCTS and performance
modeling---to a later paper.


%

\subsection{Generated Code Performance}
\label{sec:experiments:perf}
\label{sec:experiments:deadend}

We first show the code we generate compares to hand-tuned reference
implementations. We created implementation spaces for a few kernels
and compare the execution time of the best implementation found by our
exploration strategy with the reference implementation.

We created all the implementation spaces using a similar procedure. Every
instruction is placed in its own loop nest, with point to point communication between
them.
We also select a few strip-mining factors dividing the input size, for each dimension.
The search algorithm is free to reorder, fuse, unroll
or vectorize loops or to map them to the different levels of parallelism. It
can implement point-to-point communications using registers or by allocating
temporary arrays in shared memory, and chooses the  level of cache to use for each
memory accesses.

We consider the following kernels. Unless specified otherwise, the reference
implementation calls CuBLAS, Nvidia's hand-tuned implementation of basic linear
algebra kernels. Matrices are column major order.
\begin{description}
  \item[axpy]: computes $z = \alpha.x + y$, where $\alpha$ is a scalar and $x$, $y$ and
    $z$ vectors of size $n = 2^{26}$. We strip-mine $n$ twice, with factors in
    $\bbrackets{2, 4}$ and $\bbrackets{2, 1024}$.
  \item[matmul]: computes $C = A \cdot B$ when $A$ and $B$ are matrices of respective
    size $m \times k$ and $k \times n$. We strip-mine $m$ and $n$ twice, with
    factors in $\bbrackets{2, 32}$ and $\bbrackets{2, 4}$. We try different values
    for $m$, $n$ and $k$ to show how the algorithm adapts. In the rest of the paper
    we refer to them as \emph{matmul $m \times n \times k$}.
  \item[strided matmul]: is the same a matmul $1024 \times 1024 \times 1024$
    but with consecutive elements of $A$ stored with stride of 32. CuBLAS does not
    support such strides. The reference is a naive implementation
    that computes one element of $C$ per thread.
\end{description}

\begin{table*}[htp]
  \small
  \centering
  \begin{tabular}{lrrrrl}
    \toprule
    Kernel   & Space Size & Dead Ends        & Avg. Runtime & Reference   & Speedup \\
    \midrule
    axpy     & $\scinum{1.1}{11} \pm \scinum{1.7}{10}$
                          & $0.1\% \pm 0.02$ & $7.05 ms \pm 0.005$
                                                            & $10.3ms$    & \bf 1.47 $\pm 10^{-3}$ \\
    matmul $256 \times 256 \times 32$
             & $\scinum{1.83}{21} \pm \scinum{3.3}{20}$
                          & $14\% \pm 0.7$   & $34.2\mu s\pm 2.54$
                                                            & $82.8\mu s$ & \bf 2.42 $\pm 0.18$ \\
    matmul $1024 \times 1024 \times 1024$
             & $\scinum{3.5}{21} \pm \scinum{1.8}{21}$
                          & $2.8\% \pm 0.3$  & $4.81 ms \pm 0.06$
                                                            & $3.75ms$    & \bf 0.78 $\pm 0.01$ \\
    strided matmul
             & $\scinum{6.0}{20} \pm \scinum{2.0}{20}$
                          & $2.3\% \pm 0.3$  & $10.1 ms \pm 0.59$
                                                            & $637ms$     & \bf 66.7 $\pm 3.9$ \\

    \bottomrule
  \end{tabular}
  \caption{Implementation Space Exploration Results}
  \label{tab:experiments:runtimes}
\end{table*}

The benchmarks are representative of both compute-intensive (mm) and bandwidth-bound (axpy) kernels. They also span a variety of memory access patterns, including strides at different dimensions, transposed layouts,
all of these being typical of higher dimensional tensor algebra in computational chemistry, simulation codes, and machine learning \cite{TCE}.

\tabref{experiments:runtimes} summarises the characteristics of implementation spaces
and explorations results. We ran 4 explorations of 4 hours on each kernel. For
each exploration, we evaluated the best implementation 40 times. The 95\% confidence
interval on the average speedup was always within $\pm 0.5\%$.  We report the
average runtime and speedup among explorations, along with the maximum
variation of the speedup compared to the average among exploration.

The \emph{Space Size} and \emph{Dead Ends} columns respectively provide
estimations of the size of the search space (see \secref{experiments:size}) and
of the probability to encounter a dead-end when randomly walking the tree
from the root with a uniform sampling among the valid children of each node.
Dead ends occur when constraints propagation is unable to detect incompatible
decisions before more choices are specified. Both columns also indicate the
$95\%$ confidence intervals.

For all experiments, the ratio of dead-ends is inferior to a third. This shows
that finding valid implementations is easy. We only use the CSP formalism to
encode the search space and not to search for valid implementations.

\emph{axpy} and \emph{matmul $256\times256\times32$} show that our approach is
able to outperform hand-tuned implementations by finding better
implementation decisions.
\emph{matmul $1024\times1024\times1024$} compares against an implementation that almost
reaches peak performance. We have no chance of beating it as it relies on features
we do not support. In particular, it uses texture memory and manual allocation
of registers to avoid bank conflicts \cite{lai2013performance}, which is
impossible with public Nvidia APIs. However, we still achieve reasonable
performance.
\emph{strided matmul} shows the benefits of our approach for kernels not
available in hand tuned libraries, with a $66\times$ speedup over a naive
implementation. While not reaching peak performance, it is within a
factor 2.7 of CuBLAS non-strided version and thus can be useful.

The high variance on the execution time of the best implementation
among explorations on the same space is a real issue. We are working
on developing better search algorithms and refining the performance
model to mitigate the problem. This is to put in perspective with the
large size of implementation spaces and with the fact that this paper
is not about search algorithms themselves but about how to expose the
implementation space to them.

\subsection{Implementation Space Size Estimation}
\label{sec:experiments:size}

Computing the exact size of the search spaces is intractable due to the large
number of possible choices.  To estimate their size, we turn to probabilistic
methods which have been used to reliably estimate the size of search trees in
constraint solvers \cite{Kilby:2006}.

The precise method we use was described by Chen \cite{Chen:1992}, which is a
generalization of an earlier method by Knuth \cite{Knuth:1975}. The Knuth
algorithm starts at the root and performs a random descent until reaching a
leaf.  It then estimates the size of the tree by using the branching factor
along the path. The Chen algorithm, called heuristic sampling, adds the concept
of strata: classes of subtrees estimated by the algorithm designer to be
structurally similar.  The algorithm maintains a queue containing strata,
represented by a single subtree in the stratum, along with an estimate its
size.  When a subtree is discovered, the corresponding stratum in the queue is
updated with its count, and it can randomly be selected as the new
representative for the stratum.  The total size estimate is then the sum of
estimates for the leaves encountered.  A partial order on the strata which is
strictly decreasing along the tree is required to ensure well-formedness.  This
algorithm was chosen as it strikes a balance between simplicity and
performance.  We use a lexicographic pair of the depth in the tree and number
of remaining choices (assuming none get forced through propagation---this
provides a simple proxy for the subtree size to guide the algorithm) as a
stratifier.

We computed the results in \tabsref{experiments:runtimes}{experiments:sizes}
with either $1000$ iterations of the Chen method or $100,000$ iterations of the
Knuth method, whichever gave a better confidence interval. Each time, we report
the $95\%$ confidence interval. For some of the
larger implementation spaces, we perform additional iterations to bring the
confidence interval to the correct order of magnitude.

\subsection{Discriminant Information in Candidates}\label{sec:experiments:info}

We use the size estimates to justify that our representation allows extracting
pertinent information early, with only a few decisions set. \tabref{experiments:sizes}
reports the estimated size of the search space after pruning the first 10 levels
with the performance model. We cut branches whose lower bound was higher than
the average runtime obtained from the experiments of \secref{experiments:perf}.
Note that we use the execution time without the kernel launch time (obtained
using performance counters) instead of the one reported in
\tabref{experiments:runtimes} for pruning.
We also computed through exhaustive enumeration the number of nodes at depth 10
with and without cut.

These experiments show that we are able to eliminate a large portions of search
spaces early on. For most kernels, we reduce the size of the space by two or
three orders of magnitudes by cutting on the first $10$ levels.

\begin{table*}[htp]
  \small
  \centering
  \begin{tabular}{lrrrl}
    \toprule
	  Kernel   &
	  Space Size &
	  \begin{tabular}[c]{@{}c@{}}Space Size\\with cut applied\\ on first 10 levels\end{tabular} &
	  Nodes at depth 10     &
	  \begin{tabular}[c]{@{}c@{}}Nodes at depth 10\\ with cut applied\end{tabular} \\
    \midrule
    axpy &
	  $\scinum{1.1}{11} \pm \scinum{1.7}{10}$ & $\scinum{2.9}{7} \pm \scinum{1.1}{7}$ &
	  $86,587$ & $121$ ($0.14\%$)\\
    matmul $256 \times 256 \times 32$ &
	  $\scinum{1.83}{21} \pm \scinum{3.3}{20}$ & $\scinum{3.5}{18} \pm \scinum{2.2}{18}$ &
	  $134,060$ & $20,126$ ($15.0\%$) \\
    matmul $1024 \times 1024 \times 1024$ &
	  $\scinum{3.5}{21} \pm \scinum{1.8}{21}$ & $\scinum{9.5}{17} \pm \scinum{4.6}{17}$ &
	  $161,980$ & $6,738$ ($4.2\%)$\\
    strided matmul &
	  $\scinum{6.0}{20} \pm \scinum{2.0}{20}$ & $\scinum{8.1}{17} \pm
    \scinum{4.5}{17}$ & 142,780
	  & $9,512$ ($6.62\%$)\\

    \bottomrule
  \end{tabular}
  \caption{Implementation Space Exploration Results}
  \label{tab:experiments:sizes}
\end{table*}

\subsection{Impact of the Decisions Order}\label{sec:experiments:order}

One core feature of our approach is that decisions commute. Here
we show how this helps search algorithms. Experiments in
\secsref{experiments:perf}{experiments:info} first specify \texttt{memory
layout}
decisions, then \texttt{size}, then \texttt{dim\_kind},
then \texttt{thread\_mapping}, then \texttt{mem\_space}, then
\texttt{order} and last \texttt{cache}. We selected this order
to prioritise what we think are most important decisions. This allows
both the performance model and the MCTS to discriminate higher in the tree and
to focus on a fewer branches. This does not impact implementations,
only the structure of the search tree.

We ran experiments on matmul $1024\times1024\times1024$ to compare this order
with the reverse order.
In both case, we computed the ratio
of nodes that the performance model can prune in the first $T$ levels of
the tree, assuming the cut threshold is the average runtime reported in
\tabref{experiments:runtimes}. The number of candidates with a depth $\leq d$
varies greatly with the order of decisions. To have comparable results, we took
for each order the first $d$ such that the number of candidates of depth $d$ is
above $10^5$.
This resulted in $d = 10$ with $\scinum{1.6}{5}$ for the direct order
and $d = 16$ with $\scinum{1.1}{5}$ nodes for the reverse.

With the direct order, the performance model reduces the tree size by a factor
of 24. This factor falls down to $1.8$ for the reverse order. This is 13 times
more branches to consider. We also tried running the search algorithm using
the reverse order but it ran out of memory due to an explosion in the number of branches.
While we illustrated the impact of the decisions order with the
performance model, it also applies to other algorithms: picking the most
discriminant decisions first helps focusing the search.

\section{Discussion and Related Work}
\label{sec:discussion}

Traditional optimizers work on a single implementation that they iteratively
improve using rewrite rules based on pattern matching.
In constrast, we propose to work on classes of implementations modulo
optimization. We first discuss how it obviates the problem of optimization
ordering
and enable global heuristics aware of all potential optimizations. Then we
explain how our system differ from other encodings of compilation problems in
logical frameworks.

\subsection{Partial Implementations}

Ordering decisions is a common problem in compilers: transformations may enable, disable or
even undo others. Some domain-specific approaches avoid the issue with a
carefully curated set of rules. They then build and explore
a tree (or graph) whose nodes are implementations and edges rules
applications. For example, Spiral \cite{spiral} uses this approach for fast
Fourier transforms, LGen \cite{lgen} for linear algebra kernels and TCE \cite{TCE} for large tensor contractions in computational chemistry and physics simulation. More recently, LIFT
\cite{lift} applied rules to rewrite a functional representation down to low-level OpenCL
constructs and generate efficient GPU code.

However, these systems still suffer from the original problem: transformations may be
hidden behind others. In contrast, our representation allows to see from the start which
decisions are available in which branch and to make most important decisions first. An
expert programmer can even manually set decisions upfront.

An alternative to rewrite rules is to use algorithmic skeletons \cite{cole1989algorithmic}
and to map them to the hardware using a fixed strategy that leverages domain specific
information. In its simplest form, this is just parametric libraries such as
Thrust \cite{bell2011thrust} and SkelCL \cite{steuwer2011skelcl}. Otherwise, it takes the
form of high level functional operators in a domain specific language such as Delite
\cite{sujeeth2014delite}, Copperhead \cite{catanzaro2011copperhead}, Accelerate
\cite{chakravarty2011accelerating} or NOVA \cite{collins2014nova}. While theses systems
allow for optimizations, such as the fusion of operators, they rely on a fixed
strategy that makes it hard to adapt to different
hardware, different input sizes of new kind of computations.

The Sea of Nodes approach \cite{sea-of-nodes} places instructions outside of
basic blocks when possible, effectively representing classes of implementations
modulo scheduling. However, this is limited to scheduling decision.

\subsection{Global Heuristics}

The partially instantiated vector offers a complete view of potential
implementations. This allows to define heuristics aware of what a fully
optimized implementation look like. The lower bound performance model
mentioned in \secref{results} could not work if it just had access to an
intermediate implementation in the compilation process. A similar performance
model relying on ad-hoc partial implementations was previously introduced
\cite{telamon_cc17}.  We generalize the idea by encoding partial
implementations as a CSP problem on top of a semantic backbone.

Our approach is close to Equality Saturation \cite{tate2009equality}, with
similar benefits. Equality Saturation uses rewrite rules but
keeps both the original and rewritten pattern. However the number of patterns
can grow exponentially with the number of rules. In contrast,
our decision vector has a fixed size. A fixed size vector also makes it easier to
extract features for machine learning algorithms.

Frameworks that dissociate the algorithm from the schedule, such as Halide
\cite{halide} for image processing and TVM \cite{auto-tvm} for machine learning
arguably also deal with partial implementations. The algorithm is akin to our
semantics backbone and the schedule to our decision vector. However, they do not
have an easy way to reason about partial schedules.
TVM applies machine learning techniques, but only on fully
specified implementations. An interesting idea would be to use our approach
on top of their representation to explore schedules. This is also
true for other script-based transformation tools such as UTF \cite{utf} or
URUK \cite{uruk} that start from a fully specified implementation but lift it
into a mathematical representation that abstracts the initial schedule.

\subsection{Encoding as an Operation Research Problem}

The idea of encoding compilation problems in logical frameworks is not new.
Polyhedral compilation encodes transformations on loop nests as
Integer Linear Programming \cite{feautrier1,feautrier2} and PPCG \cite{ppcg} applies
it to generate code for GPUs.  Super-optimization techniques also use CSP
\cite{csp-super-opt} and SAT \cite{sat-sketching} solvers to generate optimal
sequences of instructions.

The originality of our approach is to use CSP to expose potential decisions,
not to find a solution (\secref{experiments:deadend} shows that this is easy in our case).
This allows us to manipulate whole sets of potential implementations, and is
a core reason for using CSP: it is easier to guide the search through the
domains, while SAT and ILP solvers often act more like black boxes.

ILP-based approaches maximize a single metric (such as data locality in polyhedral
schedules\cite{bondhugula2008practical}) that does not reflect the full
complexity of the architecture. 
Because we do not try to embed
performance constraints in the logical framework, we have much more flexibility
and can use a combination of custom heuristics, actual evaluations and
statistical search.

One way of solving this problem while staying in an ILP framework is to find
schedules for a single loop level at a time, starting from the outermost \cite{bondhugula2008practical}. This enables more complex heuristics and incremental evaluation at each level that goes beyond the expressive power of linear objective functions \cite{Zin18}. However, loop levels must still be considered in a fixed order. It would be interesting to use our approach with a similar encoding. Polyhedral compilers have been designed with search algorithms complementing or replacing linear programming. Pouchet et al.\ proposed custom genetic operators on affine schedules to search for dependence-preserving loop nest optimizations \cite{DBLP:conf/pldi/PouchetBCC08,pouchet2011loop}. Vasilache et al.\ constructed a two-level optimization strategy where the lower tier is a gray box based on integer linear programming, exposing a fixed set of strategies and parameters to a higher tier genetic algorithm \cite{tc}.

Diesel \cite{Elango:2018:DDL:3211346.3211354} is a recent framework from Nvidia instantiating an ILP-based scheduler for a specific domain, specializing it for each kernel with parameters specific to a GPU micro-architecture (tile sizes, mapping strategies), complemented with target-specific transformations (e.g., software pipelining, instruction-level and register-level optimizations). Code generated by Diesel reaches impressive performance levels, matching or outperforming native libraries. This involves deep knowledge of the target architecture encoded in the optimization strategy, and register-level optimizations not currently modeled in our search space.

Overall, our domain-specific language for defining decisions and constraints proved useful for developing an implementation space for linear algebra, facilitating the design and experiments with numerous decisions and constraints. We are not aware of any other approach reaching the same level of automation while remaining competitive with native GPU libraries.

\section{Conclusion}

We presented a generic approach for program optimization based on partially
specified implementations.  In our approach, optimization decisions are
available upfront and can be applied in any order, obviating the problem of
optimization ordering. The most performance-impacting decisions can be made
first, and can be specified manually through expert knowledge when available.


At the core of our representation is a decision vector listing all possible
decisions for each choice.  Global heuristics can be defined which operate on
sets of possible implementations.  They can derive pertinent information, such
as profitability estimates to guide the search and performance bounds to prune
the space, long before all decisions have been made.


Starting from a high level description of the implementation choices and their interaction, we built a tool leveraging constraint programming principles to derive efficient code for the manipulation of decision vectors. We applied this tool to the construction of a partial implementation search space for linear algebra kernels running on GPUs, generating code competitive with or outperforming carefully hand-tuned reference libraries.

So far, we limited ourselves to relatively simple search algorithms with a few remaining hardwired decisions. For example, our representation allows to search for the best decision order, and we are working on more complex algorithms that fully exploit this potential. It also enables easily sharing information between branches of the search tree as choices have fixed positions in the
decision vector, opening up more opportunities to prune the search space.


\bibliography{paper}

\end{document}